\documentclass[aps,onecolumn,groupedaddress,nofootinbib]{revtex4}
\usepackage{graphicx}
\usepackage{amsmath}
\usepackage{epsfig}

{
{

\begin{document}

\title{Origin of matter in the universe}

\author{Pei-Hong Gu}

\affiliation{The Abdus Salam International Centre for Theoretical
Physics, Strada Costiera 11, 34014 Trieste, Italy}

\begin{abstract}
We extend the standard model with two iso-singlet color triplet
scalars, one singlet real scalar and one singlet fermion. The new
fields are odd under an unbroken $Z_{2}^{}$ discrete symmetry while
the standard model particles are even. The decays of the singlet
real scalar into three standard model quarks (antiquarks) with three
singlet antifermions (fermions), which explicitly violate the baryon
number, will become effective after the electroweak phase transition
and then produce the observed baryon asymmetry in the universe
through the loop diagram involving the exchange of the $W$ gauge
boson. The singlet fermion can serve as the candidate for cold dark
matter. In our model, all new particles with masses below the TeV
scale can be detected by the forthcoming collider experiments or the
next generation experiments for neutron-antineutron oscillations.

\end{abstract}

\maketitle

\textit{Introduction}: Nowadays the baryon asymmetry in the universe
has been confirmed by the precise data from the cosmological
observations \cite{pdg2006}. This puzzle can be elegantly solved by
the leptogenesis mechanism \cite{fy1986} proposed by Fukugita and
Yanagida about twenty years ago. The salient feature of leptogenesis
is that the sphaleron processes \cite{krs1985} partially convert the
produced lepton asymmetry to the final baryon asymmetry before the
electroweak phase transition. However, the cosmological observations
\cite{pdg2006} only indicate that the present baryon asymmetry
should arise before the big bang nucleosynthesis (BBN). In other
words, any theory can solve this puzzle as long as it produces an
adequate baryon asymmetry before the BBN even if it becomes
effective after the electroweak phase transition. Recently, Babu,
Mohapatra and Nasri proposed an interesting mechanism named as
post-sphaleron baryogenesis \cite{bmn20061,bmn20062}, which need not
resort to the sphaleron processes, to explain the baryon asymmetry
in the universe.

Another major cosmological puzzle is the dark matter, which
contributes about $20\%$ to the energy density of our unverse
\cite{pdg2006}. What is the nature of dark matter? One of the poplar
candidates for dark matter is the weakly interacting massive
particles (WIMPs). Among many possible WIMPs, the lightest
superparticle (LSP) in supersymmetric models is the most widely
studied one. However, no direct experimental evidence has been
obtained for supersymmetry so that other possibilities of WIMPs,
which explain the relic density of dark matter in the universe,
should be worth studying and searching for.

In this paper, we present a nonsupersymmetric model to
simultaneously solve the puzzles of baryon asymmetry and dark matter
in the universe by extending the standard model (SM) with some new
fields. The baryon asymmetry can be produced after the electroweak
phase transition through the loop diagram involving the exchange of
the $W$ gauge boson. The relic density of cold dark matter can be
also realized as desired. In our model, all new particles with
masses below the TeV scale can be tested by the forthcoming or
proposed experiments.

\textit{The model}: We extend the $SU(3)_{c}^{}\times SU(2)_{L}^{}
\times U(1)_{Y}^{}$ SM with one singlet real scalar $A$, two
iso-singlet color triplet scalars $B,\,C$ with hypercharge
$+\frac{2}{3},\,-\frac{1}{3}$, respectively and one singlet fermion
$S$. We further introduce an unbroken $Z_{2}^{}$ discrete symmetry,
under which the new fields carry odd parity while the standard model
particles are all even. Obviously, the present model is free of
gauge anomaly. Within this framework, we have the unique baryon
number violated interaction,
\begin{eqnarray}
\label{lagrangian1} -\mathcal{L} \supset \lambda\, A\,B\,C^{2}_{}
\,+\,\textrm{h.c.}\,.
\end{eqnarray}
Furthermore, the new fields can communicate with the SM particles
through the following Yukawa couplings,
\begin{eqnarray}
\label{lagrangian2} -\mathcal{L} \supset
f_{i}^{}\,B\,\overline{u_{Ri}^{}}\,S_{}^{}
\,+\,h_{i}^{}\,C\,\overline{d_{Ri}^{}}\,S_{}^{}\,
+\,\textrm{h.c.}\,,
\end{eqnarray}
where $u_{Ri}^{}\,(\textbf{3},\,\textbf{1},\,+\frac{2}{3})$ with
$i=(u,c,t)$ is the SM up-type right-handed quark while
$d_{Ri}^{}\,(\textbf{3},\,\textbf{1},\,-\frac{1}{3})$ with
$i=(d,s,b)$ is the down-type. As for the other couplings of the new
scalar fields to the SM Higgs boson
$\phi\,(\textbf{1},\,\textbf{2},\,-\frac{1}{2})$, they can be made
negligible by choice of parameters which do not affect our
discussions.

\begin{figure} \vspace{5.5cm}
\epsfig{file=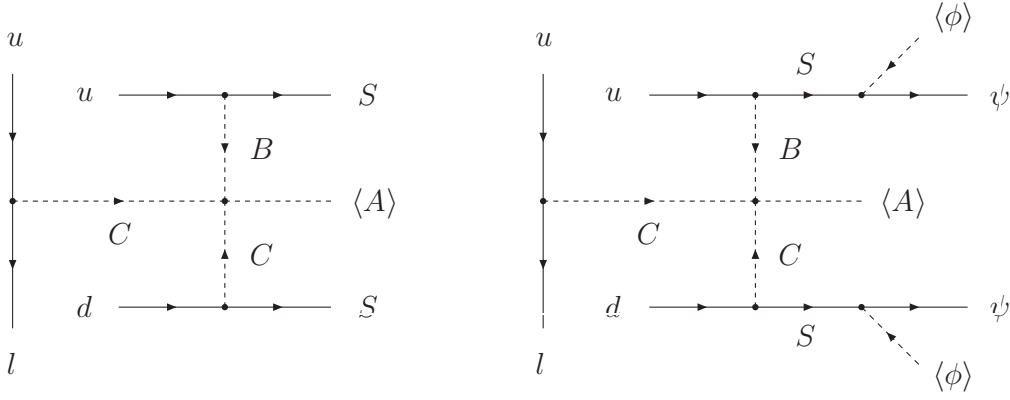, bbllx=5.2cm, bblly=7.0cm, bburx=15.2cm,
bbury=17cm, width=8.0cm, height=8.0cm, angle=0, clip=0}
\vspace{-7.5cm} \caption{\label{protondecay} The proton decay in the
case without the present $Z_{2}^{}$. The left diagram shows the
proton decay mode with the light singlet Dirac fermion while the
right is with the heavy one. Here the diagrams with the singlet
fermion being a Majorana particle have been omitted for simplicity.}
\end{figure}

Before discussing how to realize the generation of baryon asymmetry
and the relic density of dark matter, let us first clarify why the
dangerous proton decay, which appears in the usual models with the
color triplet fields \cite{bmn20061}, can be avoided in our model.
Benefited from the exact $Z_{2}^{}$, the singlet real scalar will
not develop its vacuum expectation value, the following Yukawa
interactions,
\begin{eqnarray}
\label{lagrangian3}-\mathcal{L}\supset
g^{}_{i\alpha}\,C\,\overline{u_{Ri}^{}}\,l_{R\alpha}^{c}\,+\,y_{\alpha}^{}\,\overline{\psi_{L\alpha}^{}}\,\phi
\,S_{} +\textrm{h.c.}\,,
\end{eqnarray}
will also be forbidden. Here
$\psi_{L\alpha}\,(\textbf{1},\textbf{2},-\frac{1}{2})$ and
$l_{R\alpha}^{}\,(\textbf{1},\textbf{1},-1)$ with $\alpha
=(e,\mu,\tau)$ are the SM left-handed and right-handed leptons,
respectively. Therefore, it is impossible to realize the proton
decay as shown in Fig. \ref{protondecay}, in which the proton will
either decay into one charged lepton and two singlet fermions
through a dimension-9 operator suppressed by $\frac{\langle A
\rangle}{M_{B}^{2}M_{C}^{4}}$ if the singlet fermion is light, or
decay into one charged lepton and two neutrinos through another
dimension-9 operator suppressed by
$\frac{\langle\phi\rangle^{2}_{}\langle A
\rangle}{M_{S}^{2}M_{B}^{2}M_{C}^{4}}$ if the singlet fermion is
heavy.

\begin{figure} \vspace{7.0cm}
\epsfig{file=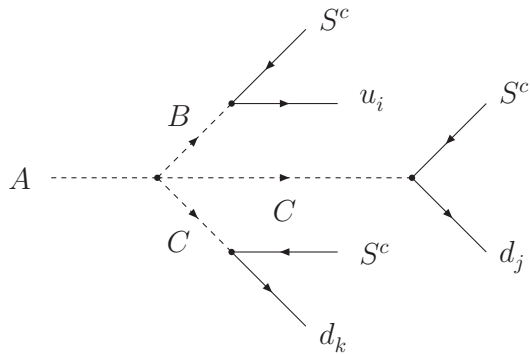, bbllx=4.5cm, bblly=7.0cm, bburx=14.5cm,
bbury=17cm, width=8.0cm, height=8.0cm, angle=0, clip=0}
\vspace{-9.0cm} \caption{\label{decay1} The singlet real scalar
field decays into three SM quarks and three singlet antifermions at
tree level. }
\end{figure}

\textit{Baryon asymmetry}: We begin to demonstrate how to generate
the observed baryon asymmetry in the universe after the electroweak
phase transition in our model. For this purpose, we assume the
following hierarchical mass spectrum for the new fields,
\begin{eqnarray}
\label{massspectrum} M_{S}^{}\sim 150 \,\textrm{GeV},\,\,
M_{A}^{}\sim 500 \,\textrm{GeV},\,\, M_{B,C}^{}\sim 600
\,\textrm{GeV}\,.
\end{eqnarray}
Therefore, the singlet real scalar $A$ can interact with three SM
(anti)quarks with the baryon number violation, $\Delta B =(-)1$, by
the exchange of the iso-singlet color triplet scalars $B,\,C$. For
example, Fig. \ref{decay1} gives the tree level process of the
singlet real scalar decaying into three quarks and three singlet
antifermions. Following \cite{bmn20061}, we can see any pre-existing
baryon asymmetry will be erased here since there are baryon number
violated interactions, which remain in equilibrium at least down to
$T_{\ast}^{}$ determined by
\begin{eqnarray}
\label{decaywidth1} \frac{1}{\left(2\pi\right)^{9}_{}}|\lambda
\tilde{f}
\tilde{h}^{2}_{}|^{2}_{}\frac{T^{13}_{\ast}}{M_{B,C}^{12}}\leq
\frac{g_{\ast}^{\frac{1}{2}}T^{2}_{\ast}}{M_{\textrm{Pl}^{}}}
\end{eqnarray}
with $\tilde{f}$ and $\tilde{h}$ being the largest of $f_{i}^{}$ and
$h_{i}^{}$. The left-handed side of the above equation is, indeed,
the temperature-dependent decay rate of the singlet real scalar. It
is straightforward to see $T_{\ast}^{}\simeq 0.2 \,M_{B,C}^{}$ with
the mass spectrum (\ref{massspectrum}) and $\lambda\sim
0.2,\,\tilde{f}\sim\tilde{h}=\mathcal{O}(1)$ by inputting the Planck
mass $M_{\textrm{Pl}}^{}\simeq 1.2\times 10^{19}_{}\,\textrm{GeV}$
as well as the relativistic degrees of freedom $g_{\ast}^{}=
\mathcal{O}(100)$ for the temperature above a few GeV.

However, as a consequence of the expansion of the universe, the
decay rate of the singlet real scalar will become a constant as soon
as the temperature falls below its mass,
\begin{eqnarray}
\label{decaywidth2} \Gamma_{A}^{} \simeq
\frac{N_{c}^{}\,P}{6^{12}_{}\times\left(2\pi\right)^{9}_{}}
|\lambda|^{2}_{}\, \textrm{Tr}\left(f^{\dagger}_{}f\right)
\left[\textrm{Tr}\left(h^{\dagger}_{}h\right)\right]^{2}_{}\frac{M^{13}_{A}}{M_{B,C}^{12}}\,,
\end{eqnarray}
where $f\equiv\left(f_{u}^{},\,f_{c}^{}\right)^{T}_{}$,
$h\equiv\left(h_{d}^{},\,h_{s}^{},\,h_{b}^{}\right)^{T}_{}$,
$N_{c}^{}=9$ is a color factor and $P\simeq 2.05$ \cite{bmn20061},
computed via Monte Carlo methods, is the phase space factor of the
six body decay. Here the top quark is absent in the decay products
due to the choice of the mass spectrum (\ref{massspectrum}). By
equating the above decay rate to the expansion rate of the universe,
\begin{eqnarray}
\label{hubble} H\simeq
1.66\,g_{\ast}^{\frac{1}{2}}\frac{T^{2}_{}}{M_{\textrm{Pl}}^{}}\,,
\end{eqnarray}
we obtain
\begin{eqnarray}
\label{temperature} T_{d}^{}\simeq\left[
\frac{N_{c}^{}\,P\,|\lambda|^{2}_{}\,
\textrm{Tr}\left(f^{\dagger}_{}f\right)
\left[\textrm{Tr}\left(h^{\dagger}_{}h\right)\right]^{2}_{}M_{\textrm{Pl}}^{}M^{13}_{A}}
{1.66\,g_{\ast}^{\frac{1}{2}}\left(2\pi\right)^{9}_{}\left(6M_{B,C}^{}\right)^{12}_{}}\right]^{\frac{1}{2}}_{}\,,
\end{eqnarray}
at which $A$ will start to decay. For instance, we deduce,
\begin{eqnarray}
\label{temperature2} T_{d}\sim 16\,\textrm{GeV}
\end{eqnarray}
with the mass spectrum (\ref{massspectrum}) as well as $\lambda\sim
0.2,\,\textrm{Tr}\left(f^{\dagger}_{}f\right)\sim
\textrm{Tr}\left(h^{\dagger}_{}h\right) = \mathcal{O}(10)$. This is
consistent with our purpose that the decay of the singlet real
scalar becomes effective before the BBN.

\begin{figure} \vspace{8.0cm}
\epsfig{file=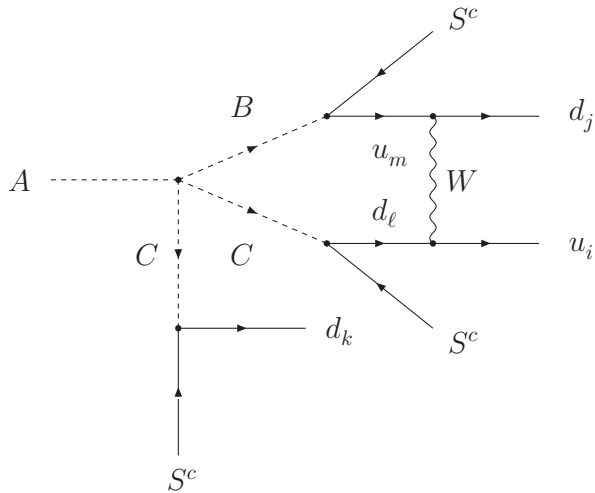, bbllx=4.5cm, bblly=7.0cm, bburx=14.5cm,
bbury=17cm, width=8.0cm, height=8.0cm, angle=0, clip=0}
\vspace{-7.0cm} \caption{\label{decay2} The singlet real scalar
field decays into three SM quarks and three singlet antifermions
through the loop diagram involving the exchange of the $W$ gauge
boson. }
\end{figure}

We now proceed to calculate the CP asymmetry which is necessary for
the dynamical generation of baryon asymmetry. Being a real scalar
field, $A$ can decay into not only three quarks with three singlet
antifermions, $A\rightarrow
u_{i}^{}d_{j}^{}d_{k}^{}S^{c}_{}S^{c}_{}S^{c}_{}$ but also three
antiquarks with three singlet fermions, $A\rightarrow
u^{c}_{i}d^{c}_{j}d^{c}_{k}SSS$. The first decay channel at tree
level has been shown in Fig. \ref{decay1}. If the CP is not
conserved, the branch ratios of the two decay channels should be
different and hence the baryon asymmetry could be expected. To
generate the CP asymmetry, we need the loop corrections to interfere
with the tree level diagram. It is definitely possible to realize
this goal by introducing other singlet real scalars. However, we
shall not adopt this approach since within the current framework an
effective loop diagram has been existing with the exchange of the
$W$ gauge boson as shown in Fig. \ref{decay2}. We derive the CP
asymmetry,
\begin{eqnarray}
\label{cpasymmetry} \varepsilon \simeq
\frac{G^{}_{F}}{\sqrt{2}\pi}\frac{\textrm{Im}\left[\textrm{Tr}\left(V^{T}_{}\hat{M}_{u}^{}ff^{\dagger}_{}
V^{\ast}_{}\hat{M}_{d}^{}hh^{\dagger}_{}
\right)\right]}{\textrm{Tr}\left(f^{\dagger}_{}f\right)
\textrm{Tr}\left(h^{\dagger}_{}h\right)}\,,
\end{eqnarray}
where $G_{F}^{}$ is the Fermi constant, $V$ is the CKM matrix,
$\hat{M}_{u}=\textrm{diag}\,(m_{u}^{},\,m_{c}^{},\,m_{t}^{})$ and
$\hat{M}_{d}=\textrm{diag}\,(m_{d}^{},\,m_{s}^{},\,m_{b}^{})$. Note
with the present choice of the mass spectrum (\ref{massspectrum}),
the contribution from the top quark should be absent in Eq.
(\ref{cpasymmetry}). We thus have
\begin{eqnarray}
\label{cpasymmetry2}\varepsilon\sim  10^{-7}\sin\delta\,
\end{eqnarray}
for $G_{F}^{}=1.17\times 10^{-5}_{}\,\textrm{GeV}^{-2}_{}$,
$V_{cb}^{}=0.04$, $m_{b}=4.20\,\textrm{GeV}$, $m_{c}^{}=1.25
\,\textrm{GeV}$, $f_{c}^{}f^{\ast}_{b}\sim h_{b}^{}h^{\ast}_{c}=
\mathcal{O}(1-10)$ and $\textrm{Tr}\left(f^{\dagger}_{}f\right)\sim
\textrm{Tr}\left(h^{\dagger}_{}h\right)= \mathcal{O}(10)$ with
$\delta$ being the CP phase.

The final baryon asymmetry can be expressed as
\begin{eqnarray}
\label{baryonasymmetry} \eta_{b}^{} \equiv
\frac{n_{b}^{}}{s}=\left(\frac{n_{b}^{}}{n_{A}^{}}\right)\left(\frac{n_{A}^{}}{s}\right)\simeq
\varepsilon \frac{T_{d}^{}}{M_{A}^{}}\,,
\end{eqnarray}
where $s=\left(2\pi^{2}_{}/45\right)\,g_{\ast}^{}\,T_{d}^{3}$ is the
entropy density and $n_{A}^{}/s$ denotes the dilution from
reheating. By using Eqs. (\ref{temperature}) and
(\ref{cpasymmetry2}), we eventually obtain
\begin{eqnarray}
\eta_{b}^{}\simeq 10^{-10}_{}
\end{eqnarray}
for the appropriate CP phase, and hence successfully explain the
observed baryon asymmetry in the universe \cite{pdg2006}.

\begin{figure} \vspace{5.0cm}
\epsfig{file=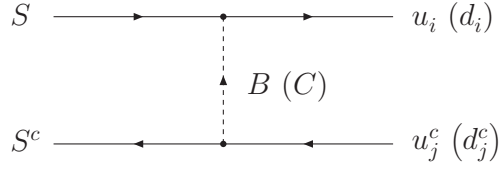, bbllx=6.5cm, bblly=7.0cm, bburx=16.5cm,
bbury=17cm, width=8.0cm, height=8.0cm, angle=0, clip=0}
\vspace{-9.5cm} \caption{\label{annihilation1} The singlet
fermion-antifermion pair annihilates into the SM quarks through the
exchange of the iso-singlet color triplet scalars. }
\end{figure}

\textit{Dark matter}: We now discuss the possibility of the singlet
fermion as the cold dark matter. Since the singlet fermion is
forbidden by the exact $Z_{2}^{}$ symmetry to have the Yukawa
couplings with the SM lepton and Higgs iso-doublets, it can not
decay into any SM particles, and become an attractive candidate for
dark matter. As shown in Fig. \ref{annihilation1}, the singlet
fermion-antifermion pair can annihilate into the SM quarks through
the exchange of the iso-singlet color triplet scalars. We have
\begin{eqnarray}
\label{crosssection2} \langle\sigma v \rangle
\simeq\frac{\left[\textrm{Tr}\left(f^{\dagger}_{}f\right)\right]^{2}_{}
+\left[\textrm{Tr}\left(h^{\dagger}_{}h\right)\right]^{2}_{}}{4\pi }
\frac{M_{S}^{2}}{M_{B,C}^{4}}\,,
\end{eqnarray}
where $\sigma$ is the total annihilation cross section of a singlet
fermion-antifermion pair, $v$ is the relative speed between $S$ and
$S^{c}_{}$ in their center-of-mass system (cms), $\langle ...
\rangle$ denotes the thermal average. Here we have used the good
approximation that for cold dark matter the average cms energy is
roughly equal to $4M_{S}^{2}$. For the mass spectrum
(\ref{massspectrum}) with
$\textrm{Tr}\left(f^{\dagger}_{}f\right)\sim
\textrm{Tr}\left(h^{\dagger}_{}h\right)= \mathcal{O}(10)$, the cross
section (\ref{crosssection2}) is about equal to $1\,\textrm{pb}$ as
would be desired to generate the right amount \cite{pdg2006} of the
relic density for cold dark matter.

\begin{figure} \vspace{5.0cm}
\epsfig{file=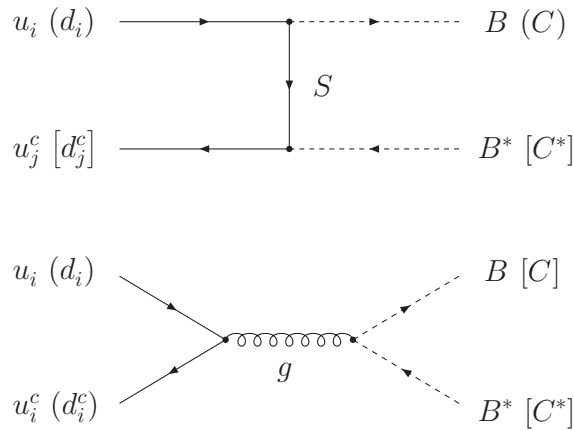, bbllx=6cm, bblly=7.0cm, bburx=16cm,
bbury=17cm, width=8.0cm, height=8.0cm, angle=0, clip=0}
\vspace{-4.0cm} \caption{\label{collider} The SM quark-antiquark
pairs annihilate into the iso-singlet color triplet scalars through
the exchange of the singlet fermion or the gluons. Here $g$ denotes
the gluons.}
\end{figure}

\begin{figure} \vspace{8.0cm}
\epsfig{file=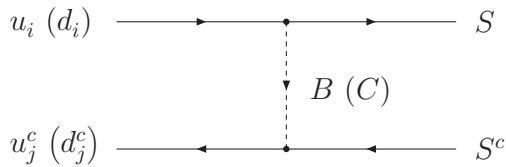, bbllx=6.5cm, bblly=7.0cm, bburx=16.5cm,
bbury=17cm, width=8.0cm, height=8.0cm, angle=0, clip=0}
\vspace{-9.5cm} \caption{\label{annihilation2} The SM
quark-antiquark pairs annihilate into the singlet fermions through
the exchange of the iso-singlet color triplet scalars. }
\end{figure}

\textit{Collider signals}: We shall expect the iso-singlet color
triplet scalars as well as the singlet fermion to be observable in
the collider experiments such as the forthcoming LHC since their
masses are below the TeV scale. As show in Fig. \ref{collider}, the
colored scalars $B,\,C$ can be produced either singly via the
processes, $u_{i}^{}+d^{c}_{j}\rightarrow
B+C^{\ast}_{},\,d_{i}^{}+u^{c}_{j}\rightarrow C+B^{\ast}_{}$, or in
pairs via the processes,
$u_{i}^{}+u^{c}_{j},\,d_{i}^{}+d^{c}_{j}\rightarrow B+B^{\ast}_{},\,
C+C^{\ast}_{}$. The two types of colored scalars can be
distinguished by their decays into the top or bottom quarks. As for
the singlet fermion, it can be produced by the annihilations of the
SM quark-antiquark pairs as shown in Fig. \ref{annihilation2}.

\begin{figure} \vspace{7.0cm}
\epsfig{file=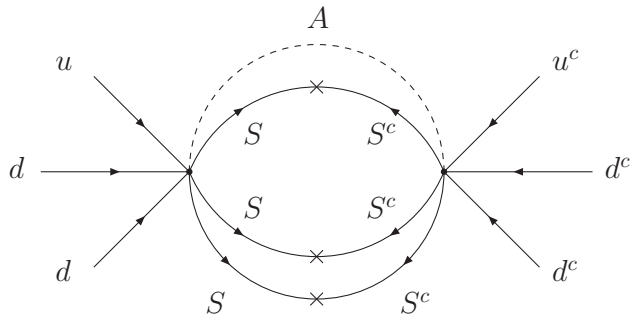, bbllx=5cm, bblly=7.0cm, bburx=15cm,
bbury=17cm, width=8.0cm, height=8.0cm, angle=0, clip=0}
\vspace{-9.5cm} \caption{\label{neutron} The neutron-antineutron
oscillation is generated by the three loop diagram if the singlet
fermion is a Majorana particle. Here the internal iso-singlet color
triplet scalars have been integrated out for simplicity.}
\end{figure}

\textit{Neutron-antineutron oscillations}: Our model can predict a
neutron-antineutron oscillation through a three loop diagram as
shown in Fig. \ref{neutron}, if the singlet fermion is a Majorana
particle. The effective strength of this neutron-antineutron
oscillation should be
\begin{eqnarray}
\label{neutronoscillation} G_{N-\bar{N}}^{} \simeq
\frac{3\left(\lambda^{\ast}_{}\right)^{2}_{} f_{u}^{2}h_{d}^{4}}
{\left(2\pi\right)^{12}}\frac{M_{S}^{9}}{M_{A}^{2}M_{B,C}^{12}}
\end{eqnarray}
by taking the cutoff at $M_{S}^{}$. For the mass spectrum
(\ref{massspectrum}) with $\lambda\sim 0.2,\,f^{2}_{u}\sim
h^{2}_{d}= \mathcal{O}(1-10)$, we have $G_{N-\bar{N}}^{}\sim
\left(10^{-30}_{}-10^{-27}_{}\right)\,\textrm{GeV}^{-5}_{}$ to be
consistent with its upper bound, $G_{N-\bar{N}}^{}\leq
10^{-27}\,\textrm{GeV}^{-5}_{}$, which corresponds to the present
limit on $\tau_{N-\bar{N}}^{}\sim 10^{8}\,\textrm{sec}$
\cite{takita1986}.

Note that the present limit on the neutron-antineutron oscillations
can be improved by two orders of magnitude in the future
\cite{kamyshkov}. It is therefore attractive that the
neutron-antineutron oscillations can be in the range accessible to
experiments and hence can be used to test our model. We should point
out that in contrast to the confirmed prediction of Refs.
\cite{bmn20061,bmn20062}, the neutron-antineutron oscillations of
our model would be absent if the singlet fermion is of Dirac nature.
Meanwhile, even if the neutron-antineutron oscillations were ruled
out in the future, our model would be still valid for solving the
puzzles of baryon asymmetry and dark matter since the singlet
fermion is free to be a Majorana or Dirac particle in the current
framework.

\textit{Conclusion}: In this paper, we extend the SM with two
iso-singlet color triplet scalars, one singlet real scalar and one
singlet fermion. The decays of the singlet real scalar into three
quarks (antiquarks) with three singlet antifermions (fermions)
through the loop diagram involving the exchange of the $W$ gauge
boson will become effective after the electroweak phase transition
and then producing the observed baryon asymmetry in the universe.
The singlet fermion-antifermion pair can annihilate into the SM
quarks and thus obtain a desired relic density to explain the puzzle
of dark matter. The neutron-antineutron oscillations will be
possible to occur in the case where the singlet fermion is a
Majorana particle. Our model can be testable at the forthcoming
experiments collider experiments or the next generation experiments
for neutron-antineutron oscillations.

\textbf{Acknowledgments}: I thank Alexei Yu. Smirnov for helpful
discussions. I also thank Rabindra Nath Mohapatra and Xinmin Zhang
for comments and suggestions.

\end{document}